\documentclass[prl,preprintnumbers,amsmath,amssymb,tightenlines,twocolumn,superscriptaddress]{revtex4-2}
\usepackage{graphicx}
\usepackage{psfrag}
\usepackage{color}
\usepackage[utf8]{inputenc}
\usepackage{ifthen}
\usepackage{amsmath}
\usepackage{listings}
\usepackage{placeins}
\usepackage{float}
\usepackage{physics}
\usepackage{comment}
\usepackage[dvipsnames]{xcolor}
\usepackage{soul}
\usepackage[inline]{enumitem}
\usepackage{braket}
\usepackage{hyperref}
\usepackage{xcolor}

\usepackage[overload]{empheq}

\usepackage{graphicx}
\usepackage{psfrag}

\usepackage{pgffor}
\usepackage{dcolumn}

\usepackage{array,ragged2e}
\usepackage[utf8]{inputenc}
\usepackage{ifthen}
\usepackage{bbold}
\usepackage{listings}
\usepackage{placeins}

\newcommand{\E}[1][\empty]{
	\ifthenelse{\equal{#1}{\empty}}
	{\mathbb{E}}
	{\mathbb{E}\left( #1 \right)}
}
\renewcommand{\exp}[1][\empty]{
	\ifthenelse{\equal{#1}{\empty}}
	{\mathrm{exp}}
	{\mathrm{e}^{#1}}
}

\newcommand{\psit}[1][\empty]{%
	\ifthenelse{\equal{#1}{\empty}}
	{\psi_t}
	{\psi_t^{(#1)}}
}

\newcommand{\npsit}[1][\empty]{%
	\ifthenelse{\equal{#1}{\empty}}
	{\tilde\psi_t}
	{\tilde\psi_t^{(#1)}}
}

\newcommand{\Bn}{\mathbf{n}}

\newcommand{\Be}{\mathbf{e}}

\definecolor{olive}{RGB}{107,142,35}

\usepackage{ulem}  
\newcommand{\oalex}[1]{{\color{blue}{}}}

\definecolor{olive}{RGB}{107,142,35}

\definecolor{orange}{RGB}{255,139,61}

\graphicspath{{{./}}}

\begin{document}

\title{Quantum Simulation of Open Quantum Dynamics via Non-Markovian Quantum State Diffusion}

\author{Yukai Guo}
\author{Xing Gao}
\email{gxing@mail.sysu.edu.cn}
\affiliation{School of Materials, Sun Yat-sen University, Shenzhen, Guangdong 518107, China }


\begin{abstract}	 
Quantum simulation of non-Markovian open quantum dynamics is essential but challenging for standard quantum computers due to their non-Hermitian nature, leading to non-unitary evolution, and the limitations of available quantum resources. Here we introduce a hybrid quantum-classical algorithm designed for simulating dissipative dynamics in system with non-Markovian environment. Our approach includes formulating a non-Markovian Stochastic Schr\"odinger equation with complex frequency modes (cNMSSE) where the non-Markovianity is characterized by the mode excitation. Following this, we utilize variational quantum simulation to capture the non-unitary evolution within the cNMSSE framework, leading to a substantial reduction in qubit requirements. To demonstrate our approach, we investigated the spin-boson model and dynamic quantum phase transitions (DQPT) within transverse field Ising model (TFIM). Significantly, our findings reveal the enhanced DQPT in TFIM due to non-Markovian behavior. 

\end{abstract}

\maketitle

An open quantum system refers to a quantum mechanical system that interacts with and is influenced by its external environment. In numerous physical and chemical processes, accounting for non-Markovianity, which involves considering the memory of the environment, becomes essential.\cite{breuer2002theory,weiss2012quantum,oliver2008charge}.
Addressing this complex scenario has prompted the development of various methodologies on classical computers~\cite{cao2014ttm,tanimura1989time,tanimura2006stochastic,yan2016deom_reivew,makri1995quapi_I,makri1995quapi_II,shi2003GQME,rabani2011GQME,mayer2000mctdh_review,wang2003multilayer,yanshao2016stochastic}.

In recent developments, significant progress has been made in enhancing the number of qubits and reducing noise rates in quantum hardware, thereby establishing a novel platform for quantum simulation~\cite{preskill2018quantum,de2021materials,arute2019quantum,daley2022practical}. 
Several quantum algorithms have been developed to simulate non-Markovian open quantum dynamics, employing diverse methodologies like the ensemble of Lindbladian trajectories\cite{head2021capturing}, the general quantum master equation\cite{wang2023simulating}, time-evolving density operator with orthogonal polynomials\cite{Guimaraes2024digitalquantum}, path integral theory\cite{walters2024path}, and dissipation equation of motion\cite{li2024towards}. These approaches primarily concentrate on evolving the reduced system density matrix. However, despite these advancements, the exploration of this domain, particularly concerning wavefunction-based quantum algorithms, remains relatively limited.
   
	\begin{figure}
		\includegraphics[width=8.6 cm]{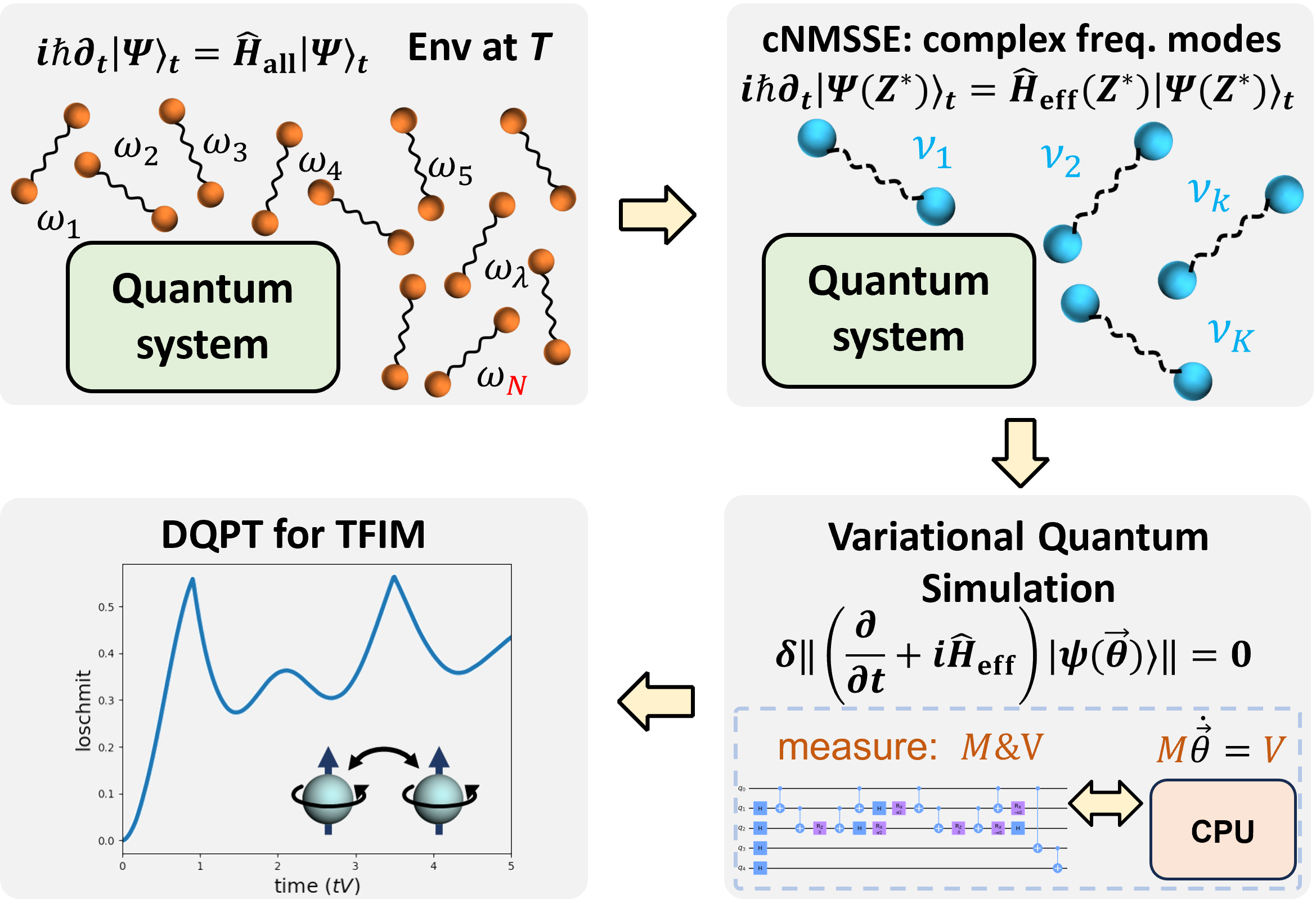}
		\caption{\label{fig1} Our strategy to simulate the non-Markovian open quantum dynamics. 
		We consider a quantum system interacting with a thermal environment at temperature $T$ consisting of bosonic modes which are linearly coupled to the quantum system. 
		After tracing out the environmental degrees of freedom, this many-body problem is then treated using the non-Markovian stochastic Schr\"odinger equation with complex frequecy modes(cNMSSE). 
		In cNMSSE, only a few modes holds significance, allowing for efficient solutions through variational quantum algorithms. This approach enables the exploration of properties like dynamical quantum phase transitions (DQPT) in the transverse field Ising model with alleviated quantum resources.}
	\end{figure}

 Concurrently, the hierarchy of stochastic pure states (HOPS)~\cite{eisfeld2014hierarchy,eisfeld2015hierarchical} method, utilizing stochastic wavefunctions, is designed for open quantum systems and aims to address the non-Markovian Quantum State Diffusion (NMQSD) equation~\cite{strunz1997pla,strunz1998pra,strunz1999prl,yu1999perturb,yu2010prl,you2014nmqsd,you2015higher}. In our prior work, we reformulated HOPS in a \textit{pseudo-Fock} space, resulting a non-Markovian stochastic Schr\"odinger equation with complex frequency modes (cNMSSE)\cite{gao2022non}. cNMSSE has been successfully solved with matrix product state and applied to study the charge transport in dissipative quantum many-body systems\cite{flannigan2022many,moroder2023stable}.  

In this Letter, we show that an efficient wavefunction-based quantum algorithm can be achieved by employing cNMSSE.
Within the framework of cNMSSE, only a few modes holds significance, enabling for efficient solutions through modern quantum algorithm, including variational quantum simulation\cite{yuan2019theory,endo2020variational} and unitary decomposition of operators\cite{schlimgen2021quantum}.
The primary emphasis in this study revolves around the application of VQS for this purpose.
We employ this approach to investigate dissipative dynamics at finite temperature within the spin-boson model, as well as dynamic quantum phase transitions (DQPT) within the transverse field Ising model (TFIM). Our findings reveal that non-Markovianity manifests through the excitation of phonon modes with complex frequencies, significantly influencing the enhancement of DQPT within TFIM.
        
Our procedure is illustrated in Fig.~\ref{fig1} and described in the following. 
	
\paragraph{The open quantum system:} 
The foundational formula of cNMSSE has been previously published in Ref.~\citenum{gao2022non}. Herein, we rearticulate the primary steps and deduce a comprehensive expression for its general form.
		We consider a quantum system coupled linearly to a (infinite) set of harmonic oscillators.
	The total Hamiltonian is written as
	\begin{equation}
		\label{eq:H_tot}
		\hat{H}_{\text{tot}}=\hat{H}_\text{S}+\hat{H}_\text{B}+\hat{H}_{\text{SB}},
	\end{equation}
	with  $\hat{H}_{\text{S}}$,  $\hat{H}_{\text{B}}$, and  $\hat{H}_{\text{SB}}$  describing the system, the bath and the system-bath interaction, respectively. 
	We consider a bath that can consist of several independent parts: $\hat{H}_\text{B}=\sum_{j=1}^J \hat{H}_{\mathrm{B},j}$ with
	$\hat{H}_{\mathrm{B},j}=\sum_\lambda (\frac{\hat{p}_{\lambda,j}^2}{2}+\frac{1}{2}\omega_{\lambda,j}^2 \hat{q}_{\lambda,j}^2)$ where  $\{\hat{p}_{\lambda,j}\}$ and $\{\hat{q}_{\lambda,j}\}$ are the coordinates and momenta of the bath DOFs.  
	The system-bath coupling Hamiltonian is taken as
	\begin{equation}
	\label{eq:H_SB}
		\hat{H}_\text{SB}=\sum_{j=1}^J H_{\text{SB},j}=\sum_j\hat{L}_j \otimes\sum_{\lambda,j} c_{\lambda,j}\hat{q}_{\lambda,j},
	\end{equation} 
	where each system operator $\hat{L}_j$ couples to its own environment.
  	 The interaction strength between system and the $(\lambda,j)$-th mode is quantified by the coefficient $c_{\lambda,j}$. 
  	 It is convenient to define the spectral densities,
  $
  		S_j(\omega)=\frac{\pi}{2}\sum_\lambda\frac{c_{\lambda,j}^2}{\omega_{\lambda,j}}\delta(\omega-\omega_{\lambda,j}),
  $
	which describes the frequency dependent system-bath coupling strength of the $j$-th bath. 
	In the time-domain, the bath correlation function,
	\begin{equation}
		\label{eq:BCF}
		\begin{aligned}
			\alpha_j(t)=&\frac{1}{\pi}\int_0^\infty \!\!\!\!\mathrm{d}\omega S_j(\omega)\big[\coth(\frac{\omega}{2 T})\cos\omega t-i\sin\omega t\big],\\
		\end{aligned}
	\end{equation}
	fully characterizes the influence of the environment at temperature $T$.
    We use the units $\hbar=k_B=1$. 
    
    We are interested in the dynamics of the system which is given by the reduced density matrix
    \begin{equation}
    	\rho(t)=\Tr_\text{B}\{{\rho_\text{tot}(t)}\}.
    \end{equation}
	Here, $\Tr_\text{B}\{\cdots\}$ denotes the trace over all bath DOFs, and $\rho_\text{tot}(t)$ is the total density matrix. 
	In the following, we assume a factorized initial state $\rho_\text{tot}(0)=\rho(0)\otimes \frac{e^{- H_\text{B}/T}}{Z_\text{B}}$ with partition function $Z_\text{B}=\Tr_\text{B}\{e^{- H_\text{B}/T}\}$.
	
	\paragraph{Non-Markovian stochastic Schr\"odinger equation and the hierarchy of pure states:}
	Within the HOPS method the reduced density operator $\rho(t)$ is obtained from
	\begin{equation}
	\label{eq:rho(t)}
	  \rho(t)=  \mathbb{E}\big\{\ket{\psi_t(Z_t^*)}\bra{\psi_t(Z_t^*)}\big\},
	\end{equation}
	where the $\ket{\psi_t(Z_t^*)}$ are vectors in the system Hilbert space that depend on  stochastic processes $Z_t$, and $\mathbb{E}[\cdots]$ denotes the average over trajectories.
	The $Z_t$ are complex valued and fulfill $\mathbb{E}[Z_t]=0$ and \footnote{With our choice of the correlation functions of the stochastic processes we follow the one of the original NMQSD derivation \cite{strunz1997pla,strunz1998pra}.	There exist other choices fore the noise-correlations which might give numerical advances e.g., for high temperature~\cite{zhao2016hierarchy,shi2016alternative}.
    }: $\mathbb{E}[Z_tZ_s]=0$ and $\mathbb{E}[Z_tZ_s^*]=\alpha(t-s)$.
	To obtain the HOPS, the bath-correlation function (\ref{eq:BCF}) is approximated by a sum of exponentials (which we denote as modes),
	\begin{equation}
		\label{eq:BCF_exp}
		\alpha(t)\approx\sum_{k=1}^{K}d_ke^{-\nu_k t}~~(t\ge 0),
	\end{equation}
	with complex numbers $\nu_k$. 
	Then the following hierarchy of first order differential equations can be derived \cite{eisfeld2014hierarchy},
	\begin{equation}
		\label{eq:hops_linear}
		\begin{aligned}
			\partial_t\psi_t^\Bn=&\big[-i\hat{H}_\text{S}+ \hat{L} Z_t^{*}-\sum_{k=1}^{K}n_k\nu_{k}\big]\psi_t^\Bn\\
			&+\hat{L} \sum_{k=1}^{K}\sqrt{d_{k}}
            \sqrt{n_k}\,\psi_t^{\Bn-\Be_k}\\
			&-\hat{L}^\dagger \sum_{k=1}^{K}\sqrt{d_{k}} \sqrt{n_k+1}\,\psi_t^{\Bn+\Be_k}.
		\end{aligned}
	\end{equation}
	The superscript $\Bn=\{n_{1},\cdots,n_{k},\cdots,n_{K}\}$ consists of a set of non-negative integer indices, and $\Be_{k}=\{0,\cdots,1_{k},\cdots,0\}$. 
	The initial conditions are $\psi^{\bf 0}_{t=0}=\psi_{\mathrm{ini}}$ and $\psi^{\Bn}_{t=0}=0$ for $\Bn\ne \mathbf{0}$.
	The trajectories entering Eq.~(\ref{eq:rho(t)}) are $\psi_t(Z^*_t)=\psi_t^\mathbf{0}(Z^*_t)$.

Note that compared to the original derivation of HOPS~\cite{eisfeld2014hierarchy} we have rescaled the auxiliary vectors according to 
$
		{\psi}^\Bn_t \rightarrow \big(\prod_{k=1}^{K}n_k!d_{k}^{n_k}\big)^{-\frac{1}{2}}\psi^\Bn_t.
$	

For the general case of several environments and several system-bath operators as given in Eq.~(\ref{eq:H_SB}), for each $\hat{L}_j$ one obtains the terms as on the right hand side of Eq.~(\ref{eq:hops_linear}), where all $k$ dependent quantities get an additional index $j$. One now has $J$ independent processes $Z^*_{t,j}$. 
The hierarchy is now labeled by $\Bn=\{n_{11},\cdots,n_{kj},\cdots,n_{KJ}\}$.
In practice one has to truncate the hierarchy, which can be achieved by a suitable approximation of the terms appearing in the last line of Eq.~(\ref{eq:hops_linear}). 
Possible choices are for example the `terminator' suggested in Ref.~\cite{eisfeld2014hierarchy}, or simply setting these terms to  to zero, as we do here.

%
	
\paragraph{Non-Markovian stochastic Schr\"odinger equation with complex frequency modes(cNMSSE):}

	To obtain at a convenient form, we formally define states
 $\{\ket{\Bn}\}$ with $\ket{\Bn}=\ket{n_{1},\cdots,n_{K}}$ and introduce
	\begin{equation}
		\label{eq:mps}
		\begin{aligned}
			\ket{\Psi(Z^*)}_t
			=&\sum_{\Bn}\psi_t^{{\Bn}}(Z^*)\ket{\Bn}
		\end{aligned}
	\end{equation}
    with the auxiliary vectors $\psi_t^{\Bn}$ of HOPS as expansion coefficients. 
   Defining the following orthonormal relation,	$\braket{\Bn|\Bn'}=\delta_{\Bn\Bn'}$, these coefficient can be obtained from $\psi_t^{{\Bn}}=\braket{\Bn|\Psi}_t$. 
The HOPS system of equations (\ref{eq:hops_linear}) is then expressed as
	\begin{equation}
		\label{eq:EOM}
		\begin{aligned}
			\partial_t\ket{\Psi(Z^*)}_t=&-i\hat{H}_{\text{eff}}(Z^*)\ket{\Psi(Z^*)}_t,\\
		\end{aligned}
	\end{equation}
	with the effective stochastic Hamiltonian
	\begin{equation}
		\label{eq:hops_h_eff_linear_rescale}
		\begin{aligned}
			\hat{H}_{\text{eff}}=&\hat{H}_\text{S}+i\hat{L} Z_t^*-i\sum_{k=1}^{K}\nu_{k}\,\hat{b}_{k}^\dagger \hat{b}_{k}\\
			&-i \sum_{k=1}^{K} \sqrt{d_{k}} (\hat{L}^\dagger\hat{b}_{k} -\hat{L} \hat{b}_{k}^\dagger),
		\end{aligned}
	\end{equation}
	where  creation $(\hat{b}^\dagger_{k})$ and annihilation $(\hat{b}_{k})$ have been defined by
	\begin{equation}
		\begin{aligned}
			\hat{b}^\dagger_{k}\ket{\Bn}=&\sqrt{n_{k}+1}\ket{\Bn+\Be_{k}}\\
			\hat{b}_{k}\ket{\Bn}=&\sqrt{n_{k}}\ket{\Bn-\Be_{k}}.
		\end{aligned}
	\end{equation}
	Now the labels $\{n_k\}$ of the hierarchy play the role of \textit{occupation numbers}. 
	Thus we refer to the states $\ket{\Bn}$ as \textit{pseudo-Fock} states.
	It is worth mentioning that the hierarchy labels $\{n_k\}$ do not appear anymore in $H_{\text{eff}}$ and that the third term in the right hand side looks like a collection of harmonic oscillators, however with complex frequencies $\{\nu_k\}$. The significant reduction in the number of modes compared to the original system offers an effective approach for quantum simulation.

By implementing the terminator $\psi_t^{n}=\frac{\hat{L}}{2\sqrt{nd}}\psi_t^{n-1}$ and considering the maximum phonon number $n_{\text{max}}=1$, the quantum state diffusion equation for a Markovian scenario can be derived.
	

\paragraph{Variational qautnum algorithm for solving cNMSSE:}
The effective Hamiltonian of Eq.~\ref{eq:hops_h_eff_linear_rescale} is essentially non-Hermitian, leading to non-unitary evolution of Eq.~\ref{eq:EOM}. Here, we employ VQS algorithm\cite{yuan2019theory,endo2020variational} to address Eq.~\ref{eq:EOM}.
VQS is a quantum-classical hybrid algorithm with broad applicability in simulating the quantum dynamics for both closed and open quantum system.
Here we adopt the Hamiltonian variational ansatz(HVA) as our variational space. Due to the non-Hermitian nature of the effective Hamiltonian given in Eq.~\ref{eq:hops_h_eff_linear_rescale}, the ansatz must be constructed with non-normalized characteristics. To this end, we decompose the effective Hamiltonian into $n$ terms $H=\sum_{i=1}^n c_iH^{(i)}$, where $c_i$ is the $i$-th coefficient and each $H^{(i)}$ corresponds to a Pauli string. 
Thus the ansatz can be represented as\cite{c1}
    $|\Phi(\Theta)\rangle = \alpha \cdot | \phi(\vec\theta))\rangle = \alpha U(\vec\theta)|\Phi_0\rangle \label{vqs3}$
with unitary gates,
    $U(\vec\theta)=\prod_{j=1}^{m} U_{n,j}(\theta_{n,j})\cdots U_{i,j}(\theta_{i,j})\cdots U_{1,j}(\theta_{1,j})~.\label{ugates}$
Here we define the parameters $\Theta:=\{\alpha, \vec\theta \}$, $\vec \theta:=\{ \theta_{i,j}\}$ and the normalized state $ | \phi(\vec\theta)\rangle :=U(\vec\theta)|\Phi_0\rangle$, where $U_{i,j}(\theta_{i,j})=e^{-iH^{(i)}\theta_{i,j}}$ represents the unitary gate controlled by the parameter $\theta_{i,j}$. The norm parameter $\alpha$ satisfies $\alpha^2 = \langle \Phi(\alpha,\vec\theta)|\Phi(\alpha,\vec\theta) \rangle$ with $\alpha  \in \mathbb{R}$. 
This ansatz consists of $m$ depths, leading to a total of $l = nm + 1$ parameters.

Instead of directly simulating the dynamics, the VQS algorithm employs McLachlan’s principle to project the original evolution onto the evolution of the parameters. This is achieved by minimizing the distance between the exact time-evolved states described by Eq.\ref{eq:EOM} and the ansatz state under infinitesimal time variation $\delta t$, which can be equivalently represented as:
\begin{align}
    \delta\|(\frac{d}{dt}+iH_\text{eff})|\Phi(\alpha,\vec\theta) \rangle \|=0~, \label{vp}
\end{align}
where $\||\Phi\rangle\|=\sqrt{\langle \Phi | \Phi \rangle}$ represents the norm of the state $|\Phi\rangle$. 
The variation is performed over the parameters($\Theta=(\alpha,\vec\theta)$), yielding an equation of motion written as\cite{c1},
\begin{align}
    \sum_{j=1}^l M_{i,j} \dot\Theta_j = V_i~, \label{le}
\end{align}
where

\begin{align}
M_{i,j}&=\begin{cases} \label{Mij}
\alpha^2 \text{Re}(\frac{\partial\langle\phi(\vec{\theta})|}{\partial \theta_i}\frac {\partial |\phi(\vec \theta)\rangle}{\partial \theta_j} )\qquad 
&i,j\neq 1
\\
1 &i,j=1
\\
0 & \text{otherwise}
\end{cases} \quad, \\ \notag
\quad V_i  &=
\begin{cases}
\alpha^2\text{Im}(\frac{\partial\langle\phi(\vec\theta)|}{\partial\theta_i} H_\text{eff}|\phi(\vec \theta)\rangle)
 &i\neq 1
\\
 \alpha \text{Im}(\langle\phi(\vec \theta)|H_\text{eff}|\phi(\vec \theta)\rangle) &i=1
\end{cases} \quad.
\end{align}
The explicit expression for the time derivative of the norm parameter, $\dot{\alpha}$, is written as:
\begin{align} 
    \dot{\alpha} &= \alpha \cdot \text{Im}\langle \phi(\vec \theta(t)) | H_\text{eff} | \phi(\vec \theta(t)) \rangle \notag
    \\ &= \alpha \cdot \frac{1}{2i}\langle \phi(\vec \theta(t)) | (H_\text{eff}-H_\text{eff}^\dagger )| \phi (\vec \theta(t)) \rangle~.\label{apara}
\end{align}

The matrix elements $M_{i,j}$  and vector element $V_i$  can be measured on quantum computers using the Hadamard test technique\cite{endo2020variational,yuan2019theory}. Subsequently, Eq.~\ref{le} can be solved on classical computer using algorithm such as the fourth order Runge-Kutta method, as in our case.

	\begin{figure}
		\includegraphics[width=8.6 cm]{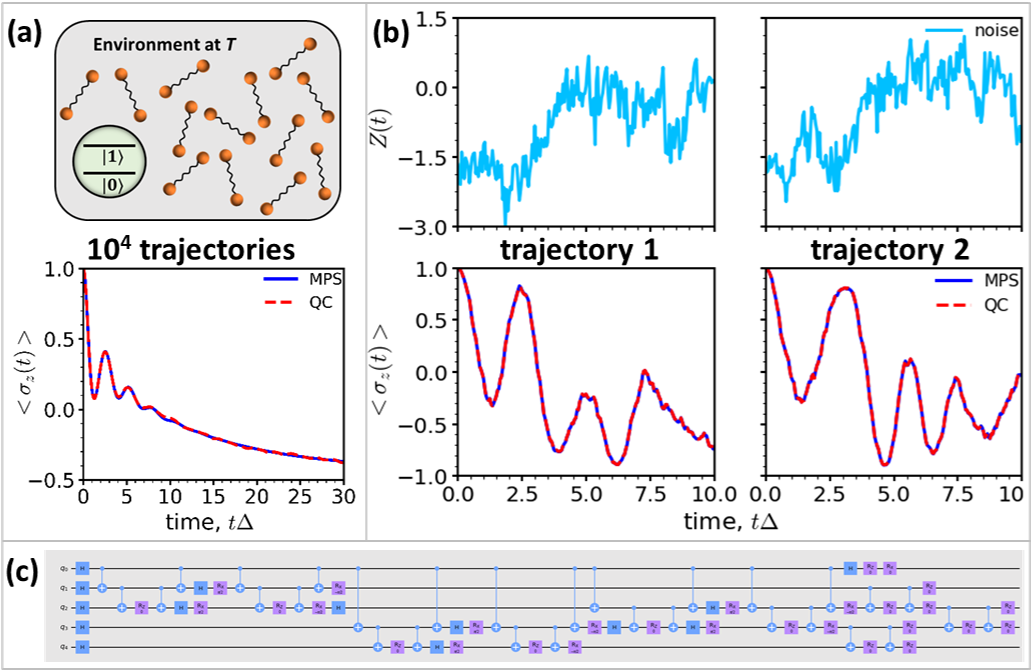}
		\caption{\label{fig2}  (a) Population dynamics of the spin-boson model with parameters $\epsilon=1.0,~\Delta=1.0,~\eta=0.5$, at a characteristic frequency $\gamma=0.25$ and temperature $\beta=0.5$, by averaging over $10^4$ (red, dashed) trajectories.
        (b) Stochastic noise and population dynamics are shown for two independent trajtories (red, dashed).
		(c) The quantum circuit diagram employed for this model. 
		The numerical exact results (blue, solid) are obtained using matrix product state method. 
		}
	\end{figure}

	\paragraph{Numerical example 1: the quantum dissipative dynamics for Spin-Boson model (SBM) at finite temperature.}
    
		The SBM is often used to test the applicability of a new method. Here $\hat{H}_\mathrm{S}=\epsilon\sigma_z+\Delta\sigma_x$ and  $\hat{L}=\sigma_z$, where  $\sigma_x=\ket{1}\bra{2}+\ket{2}\bra{1}$ and $\sigma_z=\ket{1}\bra{1}-\ket{2}\bra{2}$. 
	We consider a Debye spectral density 
	$
		S(\omega)=\eta\frac{\omega\gamma}{\omega^2+\gamma^2}.
	$
In Fig.~\ref{fig2}, we show that the population dynamics for parameters $\epsilon=1.0$, $\Delta=1.0$ and $\eta=0.5$ at finite temperature case with $\gamma=0.25$ and $\beta=0.5$. The results, both (a) averaged over $10^4$ trajectories and (b) shown for two stochastic trajectories with independent noise $Z(t)$, exhibit excellent agreement with those obtained from the matrix product state approach. 
This demonstrates the validity of our procedure.
 
Let us now consider in more detail the complexity of the equations to solve.
In each case we have chosen the number of modes $K$ large enough to guarantee convergence of the bath-correlation function.
In our case, the chosen parameters necessitate only one mode ($K=1$). Furthermore, we need to account for a maximum phonon number of $2$ for this mode. As depicted in panel (c) of Fig.~\ref{fig2}, the quantum simulation utilizes only $5$ qubits. 
\par
	
\paragraph{Numerical example 2: DQPT for TFIM}

    DQPT is a type of non-equilibrium phase transition.\cite{heyl2014dynamical,zvyagin2016dynamical,heyl2017dynamical,heyl2019dynamical} The most essential characteristic is the non-analytic behavior appearing in the characteristic functions, which is dominated by the Hamiltonian $\hat{H}$ and the initial state $\ket{\psi_{0}}$. Loschmidt echo rate function $\lambda(t)$ is usually used to describe this non-analytic behavior, which reads:
    \begin{equation}
    \label{}
        \lambda(t) = - \lim_{D \to \infty}{\frac{1}{D} \log \abs{L(t)}}
    \end{equation}
    where $D$ is the number of degrees of freedom of the system and $L(t)$ is the Loschmidt echo defined by 
    \begin{equation}
    \label{}
        L(t)= \abs{\langle\psi_{0} |e^{-i\hat{H}t}| \psi_{0}\rangle}^{2} 
            = \abs{\langle\psi_{0} | \psi_{t}\rangle}^{2}
    \end{equation}
    Owing to the size of system isn't the only criteria for observing the non-analytic behavior in the TFIM\cite{puebla2020finite,norambuena2020coding,dolgitzer2021dynamical}, we focus a system with 2 spins, where $\hat{H}_\mathrm{S}= -J\hat{\sigma}_{1}^{x}\hat{\sigma}_{2}^{x} - B(\hat{\sigma}_{1}^{z} + \hat{\sigma}_{2}^{z}) $ and $\hat{L}= \sum_{i=1}^{2} \hat{\sigma}_{i}^{+} = \sum_{i=1}^{2} \frac{1}{2} (\hat{\sigma}_{i}^{x}+ i\hat{\sigma}_{i}^{y})$.
    For observing the impact of the Markovian and non-Markovian environments, we adopt the Ornstein-Uhlenbeck type correlation function:
    \begin{equation}
    \label{}
        \alpha(t-s)= \frac{\Gamma \gamma}{2} e^{-\gamma \abs{t-s}}
    \end{equation}
    Here, $\Gamma$ is the coupling strength between the system and the environment, $\frac{1}{\gamma}$ is the memory time of the environment. When $\gamma$ changes from $\infty$ to a finite number, the environment is transferred from the Markovian to the non-Markovian regime.  Obviously, with the same form as Eq.~(\ref{eq:BCF_exp}), one mode($K=1$) is enough to describe the environment. To study the DQPT of this system, we introduce the following rate function $\Lambda(t)$:
    \begin{equation}
    \label{}
        \Lambda(t)= \min_{i=0,1}{(-\frac{1}{2}\ln{\abs{\langle\psi_{i} | \psi_{t}\rangle}^{2}})}
    \end{equation}
    where the initial state is prepared as $\ket{\psi_{0}}= \frac{1}{2} (\ket{\uparrow\uparrow} +\ket{\uparrow\downarrow}+\ket{\downarrow\uparrow}+\ket{\downarrow\downarrow})$, and $\ket{\psi_{1}} = (\ket{\uparrow\uparrow} -\ket{\uparrow\downarrow} - \ket{\downarrow\uparrow}+\ket{\downarrow\downarrow})$ is the other degenerate ground state in terms of $\hat{-\sigma_1^x\sigma_2^x}$. Prior to presenting the results, a brief description of relevant parameters of model is warranted as below: $J=2$, $B=1/0.42$ and $\Gamma=1$.

	\begin{figure}
		\includegraphics[width=8.6cm]{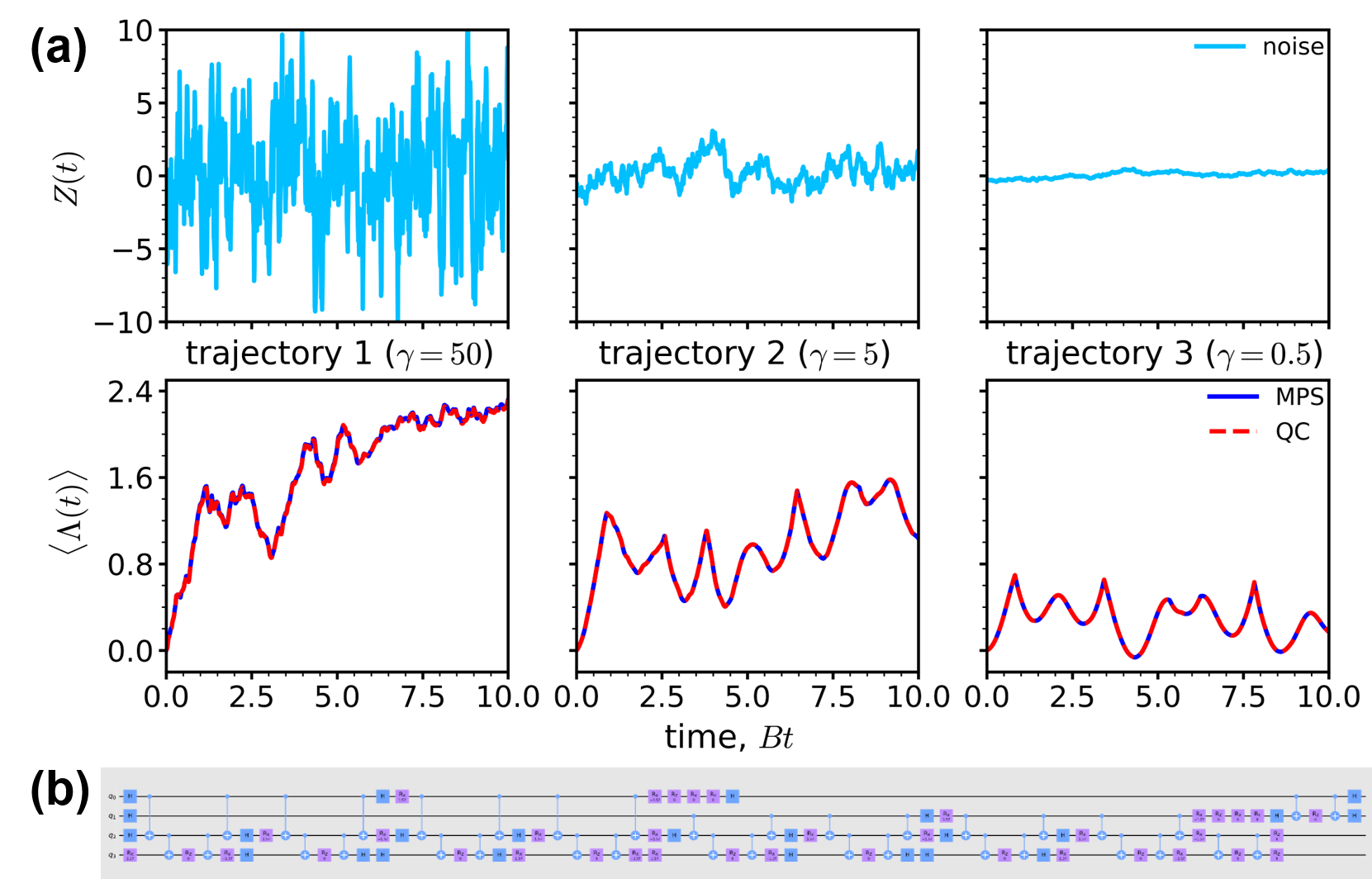}
		\caption{\label{fig3} (a) Stochastic noise and rate function $\Lambda(t)$ for one trajectory of different $\gamma$ (which is set to $50$, $5$, $0.5$ from left to right), other parameters of TFIM: $J=2$, $B=1/0.42$, and $\Gamma=1$(red, dashed).
	    (b) The quantum circuit diagram employed for this model. 
		The numerical exact results (blue, solid) are obtained using matrix product state method. }
	\end{figure}

\begin{figure}[htbp!]
    \includegraphics[width=8.6cm]{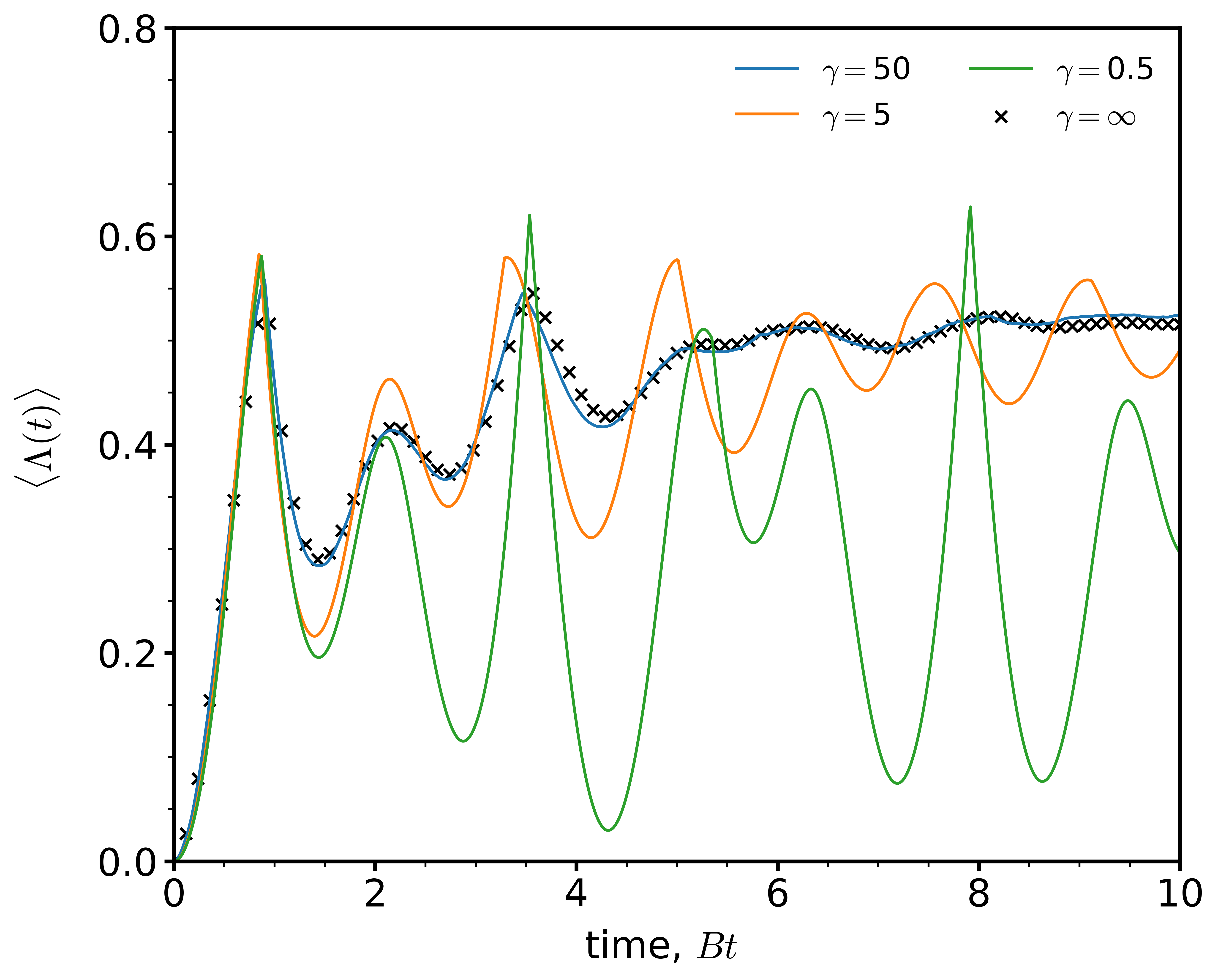}
    \caption{\label{fig4} rate function $\Lambda(t)$ of transverse field Ising model with parameters $J=2$, $B=1/0.42$ and $\Gamma=1$. In the Markovian range($\gamma=50$), the result of quantum simulation(solid line) coincides with that of Lindblad master equation calculated by QuTiP(crosses). When decreasing $\gamma$, the environment transitions to the non-Markovian regime, resulting in more occurrences of DQPT. 
    }
\end{figure}

    In Fig.~\ref{fig3}, we show the rate function of one trajectory for different $\gamma$ with $Z(t)$ generated by the solution of Langevin stochastic differential equation in \cite{gao2019charge}. For larger $\gamma$, the oscillations of $Z(t)$ become more pronounced, indicating a transition of the stochastic term in (\ref{eq:hops_h_eff_linear_rescale}) from non-Markovian to Markovian environment. Compared with matrix product state method, our method produced the consistent results. Since only one mode($K=1$) is required, the number of qubits is solely dependent on the maximum phonon number and encoding method of bosons\cite{somma2003quantum,sawaya2020resource,di2021improving}. Here, for the case of $\gamma=50$, only a maximum phonon number of 1 is required, which corresponds to the Markovian scenario (as evident from Fig.~\ref{fig4}). And a maximum phonon number of 2 is needed for other cases. For the mapping to the qubit basis states, we choose the unary ("one-hot") encoding for bosons, leading to 4 and 5 qubits needed for quantum simulation for above two cases. 

    In Fig.~\ref{fig4}, we demonstrate the rate function for different $\gamma$, which is averaged over $10^4$ trajectories. It should be emphasized that, when $\gamma=50$, the result obtained by the quantum simulation coincides with that computed using QuTiP \cite{johansson2012qutip,JOHANSSON2013qutip} for the Lindblad master equation, indicating the environment is Markovian. When decreasing $\gamma$, the environment transitions to the non-Markovian regime. In the Markovian regime, DQPT only occurs at the begining, indicating the system evolves to a steady state quickly. While for the non-Markovian environment, the occurrence of DQPT increases as $\gamma$ decreases. It can be observed that for $\gamma=0.5$, the oscillation period is close to the memory time of the environment.

	\paragraph{Conclusions:}
	\label{sec:summary}
	
	In summary, this study presents the development of a cNMSSE-VQS algorithm designed to simulate open quantum dynamics at finite temperature, accommodating both Markovian and non-Markovian environments. The algorithm employs stochastic unraveling of density matrix evolution using cNMSSE, followed by VQS implementation. Validation of this approach was conducted by examining dissipative dynamics in the spin-boson model and DQPT in the four-level TFIM. Our findings demonstrate the algorithm's high accuracy and efficiency, as evidenced by its robustness and convergence across various parameter settings within these models. Furthermore, we observe enhanced DQPT effects with increased environmental memory. Notably, compared to density matrix-based methods, cNMSSE operates on pure states, thereby reducing the qubit requirements for encoding. This characteristic makes it particularly well-suited for NISQ devices.

    The cNMSSE-VQS algorithm holds potential for further extension and broader developments. 
    Here, alternative advanced quantum algorithms,like the general unitary decomposition of operators approach\cite{qs_unitary_decomposition}, can substitute the VQS algorithm to manage non-Hermitian Hamiltonians. Additionally, the phonon space within cNMSSE can be optimized using variational techniques\cite{li2023efficient}.
    The bath correlation function can be generalized beyond a mere approximation as a sum of exponentials, allowing for non-exponential expansions\cite{ikeda2020generalization}. 
    The successful applications of the NMSSE to the Holstein-Hubbard model\cite{moroder2023stable} have inspired us to delve deeper into dissipative dynamics within many-body open quantum systems, driven by advancements in cNMSSE based quantum simulation techniques.

	\vspace{0.9cm}

	\begin{acknowledgements}
	    X.G.~acknowledge support from National Natural Science Foundation of China(Grant No. 22273122,T2350009), the Guangdong Provincial Natural Science Foundation (Grant No. 2024A1515011504), and computational resources and services provided by national supercomputer center in Guangzhou. 
	\end{acknowledgements}

	\bibliographystyle{apsrev4-2}
	\bibliography{ref_homps}

\begin{thebibliography}{58}%
\makeatletter
\providecommand \@ifxundefined [1]{%
 \@ifx{#1\undefined}
}%
\providecommand \@ifnum [1]{%
 \ifnum #1\expandafter \@firstoftwo
 \else \expandafter \@secondoftwo
 \fi
}%
\providecommand \@ifx [1]{%
 \ifx #1\expandafter \@firstoftwo
 \else \expandafter \@secondoftwo
 \fi
}%
\providecommand \natexlab [1]{#1}%
\providecommand \enquote  [1]{``#1''}%
\providecommand \bibnamefont  [1]{#1}%
\providecommand \bibfnamefont [1]{#1}%
\providecommand \citenamefont [1]{#1}%
\providecommand \href@noop [0]{\@secondoftwo}%
\providecommand \href [0]{\begingroup \@sanitize@url \@href}%
\providecommand \@href[1]{\@@startlink{#1}\@@href}%
\providecommand \@@href[1]{\endgroup#1\@@endlink}%
\providecommand \@sanitize@url [0]{\catcode `\\12\catcode `\$12\catcode `\&12\catcode `\#12\catcode `\^12\catcode `\_12\catcode `\%12\relax}%
\providecommand \@@startlink[1]{}%
\providecommand \@@endlink[0]{}%
\providecommand \url  [0]{\begingroup\@sanitize@url \@url }%
\providecommand \@url [1]{\endgroup\@href {#1}{\urlprefix }}%
\providecommand \urlprefix  [0]{URL }%
\providecommand \Eprint [0]{\href }%
\providecommand \doibase [0]{https://doi.org/}%
\providecommand \selectlanguage [0]{\@gobble}%
\providecommand \bibinfo  [0]{\@secondoftwo}%
\providecommand \bibfield  [0]{\@secondoftwo}%
\providecommand \translation [1]{[#1]}%
\providecommand \BibitemOpen [0]{}%
\providecommand \bibitemStop [0]{}%
\providecommand \bibitemNoStop [0]{.\EOS\space}%
\providecommand \EOS [0]{\spacefactor3000\relax}%
\providecommand \BibitemShut  [1]{\csname bibitem#1\endcsname}%
\let\auto@bib@innerbib\@empty
\bibitem [{\citenamefont {Breuer}\ and\ \citenamefont {Petruccione}(2002)}]{breuer2002theory}%
  \BibitemOpen
  \bibfield  {author} {\bibinfo {author} {\bibfnamefont {H.-P.}\ \bibnamefont {Breuer}}\ and\ \bibinfo {author} {\bibfnamefont {F.}~\bibnamefont {Petruccione}},\ }\href@noop {} {\emph {\bibinfo {title} {The Theory of Open Quantum Systems}}}\ (\bibinfo  {publisher} {Oxford University Press},\ \bibinfo {year} {2002})\BibitemShut {NoStop}%
\bibitem [{\citenamefont {Weiss}(2012)}]{weiss2012quantum}%
  \BibitemOpen
  \bibfield  {author} {\bibinfo {author} {\bibfnamefont {U.}~\bibnamefont {Weiss}},\ }\href@noop {} {\emph {\bibinfo {title} {Quantum dissipative systems}}},\ Vol.~\bibinfo {volume} {13}\ (\bibinfo  {publisher} {World scientific},\ \bibinfo {year} {2012})\BibitemShut {NoStop}%
\bibitem [{\citenamefont {May}\ and\ \citenamefont {K{\"u}hn}(2008)}]{oliver2008charge}%
  \BibitemOpen
  \bibfield  {author} {\bibinfo {author} {\bibfnamefont {V.}~\bibnamefont {May}}\ and\ \bibinfo {author} {\bibfnamefont {O.}~\bibnamefont {K{\"u}hn}},\ }\href@noop {} {\emph {\bibinfo {title} {Charge and Energy Transfer Dynamics in Molecular Systems}}}\ (\bibinfo  {publisher} {John Wiley \& Sons},\ \bibinfo {year} {2008})\BibitemShut {NoStop}%
\bibitem [{\citenamefont {Cerrillo}\ and\ \citenamefont {Cao}(2014)}]{cao2014ttm}%
  \BibitemOpen
  \bibfield  {author} {\bibinfo {author} {\bibfnamefont {J.}~\bibnamefont {Cerrillo}}\ and\ \bibinfo {author} {\bibfnamefont {J.}~\bibnamefont {Cao}},\ }\href@noop {} {\bibfield  {journal} {\bibinfo  {journal} {Phys.~Rev.~Lett.}\ }\textbf {\bibinfo {volume} {112}},\ \bibinfo {pages} {110401} (\bibinfo {year} {2014})}\BibitemShut {NoStop}%
\bibitem [{\citenamefont {Tanimura}\ and\ \citenamefont {Kubo}(1989)}]{tanimura1989time}%
  \BibitemOpen
  \bibfield  {author} {\bibinfo {author} {\bibfnamefont {Y.}~\bibnamefont {Tanimura}}\ and\ \bibinfo {author} {\bibfnamefont {R.}~\bibnamefont {Kubo}},\ }\href@noop {} {\bibfield  {journal} {\bibinfo  {journal} {J. Phys. Soc. Jpn.}\ }\textbf {\bibinfo {volume} {58}},\ \bibinfo {pages} {101} (\bibinfo {year} {1989})}\BibitemShut {NoStop}%
\bibitem [{\citenamefont {Tanimura}(2006)}]{tanimura2006stochastic}%
  \BibitemOpen
  \bibfield  {author} {\bibinfo {author} {\bibfnamefont {Y.}~\bibnamefont {Tanimura}},\ }\href@noop {} {\bibfield  {journal} {\bibinfo  {journal} {J. Phys. Soc. Jpn.}\ }\textbf {\bibinfo {volume} {75}},\ \bibinfo {pages} {082001} (\bibinfo {year} {2006})}\BibitemShut {NoStop}%
\bibitem [{\citenamefont {Yan}\ \emph {et~al.}(2016)\citenamefont {Yan}, \citenamefont {Jin}, \citenamefont {Xu},\ and\ \citenamefont {Zheng}}]{yan2016deom_reivew}%
  \BibitemOpen
  \bibfield  {author} {\bibinfo {author} {\bibfnamefont {Y.}~\bibnamefont {Yan}}, \bibinfo {author} {\bibfnamefont {J.}~\bibnamefont {Jin}}, \bibinfo {author} {\bibfnamefont {R.-X.}\ \bibnamefont {Xu}},\ and\ \bibinfo {author} {\bibfnamefont {X.}~\bibnamefont {Zheng}},\ }\href@noop {} {\bibfield  {journal} {\bibinfo  {journal} {Front. Phys.}\ }\textbf {\bibinfo {volume} {11}},\ \bibinfo {pages} {110306} (\bibinfo {year} {2016})}\BibitemShut {NoStop}%
\bibitem [{\citenamefont {Makri}\ and\ \citenamefont {Makarov}(1995{\natexlab{a}})}]{makri1995quapi_I}%
  \BibitemOpen
  \bibfield  {author} {\bibinfo {author} {\bibfnamefont {N.}~\bibnamefont {Makri}}\ and\ \bibinfo {author} {\bibfnamefont {D.~E.}\ \bibnamefont {Makarov}},\ }\href@noop {} {\bibfield  {journal} {\bibinfo  {journal} {J.~Chem.~Phys.}\ }\textbf {\bibinfo {volume} {102}},\ \bibinfo {pages} {4600} (\bibinfo {year} {1995}{\natexlab{a}})}\BibitemShut {NoStop}%
\bibitem [{\citenamefont {Makri}\ and\ \citenamefont {Makarov}(1995{\natexlab{b}})}]{makri1995quapi_II}%
  \BibitemOpen
  \bibfield  {author} {\bibinfo {author} {\bibfnamefont {N.}~\bibnamefont {Makri}}\ and\ \bibinfo {author} {\bibfnamefont {D.~E.}\ \bibnamefont {Makarov}},\ }\href@noop {} {\bibfield  {journal} {\bibinfo  {journal} {J.~Chem.~Phys.}\ }\textbf {\bibinfo {volume} {102}},\ \bibinfo {pages} {4611} (\bibinfo {year} {1995}{\natexlab{b}})}\BibitemShut {NoStop}%
\bibitem [{\citenamefont {Shi}\ and\ \citenamefont {Geva}(2003)}]{shi2003GQME}%
  \BibitemOpen
  \bibfield  {author} {\bibinfo {author} {\bibfnamefont {Q.}~\bibnamefont {Shi}}\ and\ \bibinfo {author} {\bibfnamefont {E.}~\bibnamefont {Geva}},\ }\href@noop {} {\bibfield  {journal} {\bibinfo  {journal} {J.~Chem.~Phys.}\ }\textbf {\bibinfo {volume} {119}},\ \bibinfo {pages} {12063} (\bibinfo {year} {2003})}\BibitemShut {NoStop}%
\bibitem [{\citenamefont {Cohen}\ and\ \citenamefont {Rabani}(2011)}]{rabani2011GQME}%
  \BibitemOpen
  \bibfield  {author} {\bibinfo {author} {\bibfnamefont {G.}~\bibnamefont {Cohen}}\ and\ \bibinfo {author} {\bibfnamefont {E.}~\bibnamefont {Rabani}},\ }\href@noop {} {\bibfield  {journal} {\bibinfo  {journal} {Phys.~Rev.~B}\ }\textbf {\bibinfo {volume} {84}},\ \bibinfo {pages} {075150} (\bibinfo {year} {2011})}\BibitemShut {NoStop}%
\bibitem [{\citenamefont {Beck}\ \emph {et~al.}(2000)\citenamefont {Beck}, \citenamefont {J{\"a}ckle}, \citenamefont {Worth},\ and\ \citenamefont {Meyer}}]{mayer2000mctdh_review}%
  \BibitemOpen
  \bibfield  {author} {\bibinfo {author} {\bibfnamefont {M.~H.}\ \bibnamefont {Beck}}, \bibinfo {author} {\bibfnamefont {A.}~\bibnamefont {J{\"a}ckle}}, \bibinfo {author} {\bibfnamefont {G.~A.}\ \bibnamefont {Worth}},\ and\ \bibinfo {author} {\bibfnamefont {H.-D.}\ \bibnamefont {Meyer}},\ }\href@noop {} {\bibfield  {journal} {\bibinfo  {journal} {Phys. Rep.}\ }\textbf {\bibinfo {volume} {324}},\ \bibinfo {pages} {1} (\bibinfo {year} {2000})}\BibitemShut {NoStop}%
\bibitem [{\citenamefont {Wang}\ and\ \citenamefont {Thoss}(2003)}]{wang2003multilayer}%
  \BibitemOpen
  \bibfield  {author} {\bibinfo {author} {\bibfnamefont {H.}~\bibnamefont {Wang}}\ and\ \bibinfo {author} {\bibfnamefont {M.}~\bibnamefont {Thoss}},\ }\href@noop {} {\bibfield  {journal} {\bibinfo  {journal} {J.~Chem.~Phys.}\ }\textbf {\bibinfo {volume} {119}},\ \bibinfo {pages} {1289} (\bibinfo {year} {2003})}\BibitemShut {NoStop}%
\bibitem [{\citenamefont {Yan}\ and\ \citenamefont {Shao}(2016)}]{yanshao2016stochastic}%
  \BibitemOpen
  \bibfield  {author} {\bibinfo {author} {\bibfnamefont {Y.-A.}\ \bibnamefont {Yan}}\ and\ \bibinfo {author} {\bibfnamefont {J.}~\bibnamefont {Shao}},\ }\href@noop {} {\bibfield  {journal} {\bibinfo  {journal} {Frontiers of Physics}\ }\textbf {\bibinfo {volume} {11}},\ \bibinfo {pages} {1} (\bibinfo {year} {2016})}\BibitemShut {NoStop}%
\bibitem [{\citenamefont {Preskill}(2018)}]{preskill2018quantum}%
  \BibitemOpen
  \bibfield  {author} {\bibinfo {author} {\bibfnamefont {J.}~\bibnamefont {Preskill}},\ }\href@noop {} {\bibfield  {journal} {\bibinfo  {journal} {Quantum}\ }\textbf {\bibinfo {volume} {2}},\ \bibinfo {pages} {79} (\bibinfo {year} {2018})}\BibitemShut {NoStop}%
\bibitem [{\citenamefont {De~Leon}\ \emph {et~al.}(2021)\citenamefont {De~Leon}, \citenamefont {Itoh}, \citenamefont {Kim}, \citenamefont {Mehta}, \citenamefont {Northup}, \citenamefont {Paik}, \citenamefont {Palmer}, \citenamefont {Samarth}, \citenamefont {Sangtawesin},\ and\ \citenamefont {Steuerman}}]{de2021materials}%
  \BibitemOpen
  \bibfield  {author} {\bibinfo {author} {\bibfnamefont {N.~P.}\ \bibnamefont {De~Leon}}, \bibinfo {author} {\bibfnamefont {K.~M.}\ \bibnamefont {Itoh}}, \bibinfo {author} {\bibfnamefont {D.}~\bibnamefont {Kim}}, \bibinfo {author} {\bibfnamefont {K.~K.}\ \bibnamefont {Mehta}}, \bibinfo {author} {\bibfnamefont {T.~E.}\ \bibnamefont {Northup}}, \bibinfo {author} {\bibfnamefont {H.}~\bibnamefont {Paik}}, \bibinfo {author} {\bibfnamefont {B.}~\bibnamefont {Palmer}}, \bibinfo {author} {\bibfnamefont {N.}~\bibnamefont {Samarth}}, \bibinfo {author} {\bibfnamefont {S.}~\bibnamefont {Sangtawesin}},\ and\ \bibinfo {author} {\bibfnamefont {D.~W.}\ \bibnamefont {Steuerman}},\ }\href@noop {} {\bibfield  {journal} {\bibinfo  {journal} {Science}\ }\textbf {\bibinfo {volume} {372}},\ \bibinfo {pages} {eabb2823} (\bibinfo {year} {2021})}\BibitemShut {NoStop}%
\bibitem [{\citenamefont {Arute}\ \emph {et~al.}(2019)\citenamefont {Arute}, \citenamefont {Arya}, \citenamefont {Babbush}, \citenamefont {Bacon}, \citenamefont {Bardin}, \citenamefont {Barends}, \citenamefont {Biswas}, \citenamefont {Boixo}, \citenamefont {Brandao}, \citenamefont {Buell} \emph {et~al.}}]{arute2019quantum}%
  \BibitemOpen
  \bibfield  {author} {\bibinfo {author} {\bibfnamefont {F.}~\bibnamefont {Arute}}, \bibinfo {author} {\bibfnamefont {K.}~\bibnamefont {Arya}}, \bibinfo {author} {\bibfnamefont {R.}~\bibnamefont {Babbush}}, \bibinfo {author} {\bibfnamefont {D.}~\bibnamefont {Bacon}}, \bibinfo {author} {\bibfnamefont {J.~C.}\ \bibnamefont {Bardin}}, \bibinfo {author} {\bibfnamefont {R.}~\bibnamefont {Barends}}, \bibinfo {author} {\bibfnamefont {R.}~\bibnamefont {Biswas}}, \bibinfo {author} {\bibfnamefont {S.}~\bibnamefont {Boixo}}, \bibinfo {author} {\bibfnamefont {F.~G.}\ \bibnamefont {Brandao}}, \bibinfo {author} {\bibfnamefont {D.~A.}\ \bibnamefont {Buell}}, \emph {et~al.},\ }\href@noop {} {\bibfield  {journal} {\bibinfo  {journal} {Nature}\ }\textbf {\bibinfo {volume} {574}},\ \bibinfo {pages} {505} (\bibinfo {year} {2019})}\BibitemShut {NoStop}%
\bibitem [{\citenamefont {Daley}\ \emph {et~al.}(2022)\citenamefont {Daley}, \citenamefont {Bloch}, \citenamefont {Kokail}, \citenamefont {Flannigan}, \citenamefont {Pearson}, \citenamefont {Troyer},\ and\ \citenamefont {Zoller}}]{daley2022practical}%
  \BibitemOpen
  \bibfield  {author} {\bibinfo {author} {\bibfnamefont {A.~J.}\ \bibnamefont {Daley}}, \bibinfo {author} {\bibfnamefont {I.}~\bibnamefont {Bloch}}, \bibinfo {author} {\bibfnamefont {C.}~\bibnamefont {Kokail}}, \bibinfo {author} {\bibfnamefont {S.}~\bibnamefont {Flannigan}}, \bibinfo {author} {\bibfnamefont {N.}~\bibnamefont {Pearson}}, \bibinfo {author} {\bibfnamefont {M.}~\bibnamefont {Troyer}},\ and\ \bibinfo {author} {\bibfnamefont {P.}~\bibnamefont {Zoller}},\ }\href@noop {} {\bibfield  {journal} {\bibinfo  {journal} {Nature}\ }\textbf {\bibinfo {volume} {607}},\ \bibinfo {pages} {667} (\bibinfo {year} {2022})}\BibitemShut {NoStop}%
\bibitem [{\citenamefont {Head-Marsden}\ \emph {et~al.}(2021)\citenamefont {Head-Marsden}, \citenamefont {Krastanov}, \citenamefont {Mazziotti},\ and\ \citenamefont {Narang}}]{head2021capturing}%
  \BibitemOpen
  \bibfield  {author} {\bibinfo {author} {\bibfnamefont {K.}~\bibnamefont {Head-Marsden}}, \bibinfo {author} {\bibfnamefont {S.}~\bibnamefont {Krastanov}}, \bibinfo {author} {\bibfnamefont {D.~A.}\ \bibnamefont {Mazziotti}},\ and\ \bibinfo {author} {\bibfnamefont {P.}~\bibnamefont {Narang}},\ }\href@noop {} {\bibfield  {journal} {\bibinfo  {journal} {Phys.~Rev.~Research.}\ }\textbf {\bibinfo {volume} {3}},\ \bibinfo {pages} {013182} (\bibinfo {year} {2021})}\BibitemShut {NoStop}%
\bibitem [{\citenamefont {Wang}\ \emph {et~al.}(2023)\citenamefont {Wang}, \citenamefont {Mulvihill}, \citenamefont {Hu}, \citenamefont {Lyu}, \citenamefont {Shivpuje}, \citenamefont {Liu}, \citenamefont {Soley}, \citenamefont {Geva}, \citenamefont {Batista},\ and\ \citenamefont {Kais}}]{wang2023simulating}%
  \BibitemOpen
  \bibfield  {author} {\bibinfo {author} {\bibfnamefont {Y.}~\bibnamefont {Wang}}, \bibinfo {author} {\bibfnamefont {E.}~\bibnamefont {Mulvihill}}, \bibinfo {author} {\bibfnamefont {Z.}~\bibnamefont {Hu}}, \bibinfo {author} {\bibfnamefont {N.}~\bibnamefont {Lyu}}, \bibinfo {author} {\bibfnamefont {S.}~\bibnamefont {Shivpuje}}, \bibinfo {author} {\bibfnamefont {Y.}~\bibnamefont {Liu}}, \bibinfo {author} {\bibfnamefont {M.~B.}\ \bibnamefont {Soley}}, \bibinfo {author} {\bibfnamefont {E.}~\bibnamefont {Geva}}, \bibinfo {author} {\bibfnamefont {V.~S.}\ \bibnamefont {Batista}},\ and\ \bibinfo {author} {\bibfnamefont {S.}~\bibnamefont {Kais}},\ }\href@noop {} {\bibfield  {journal} {\bibinfo  {journal} {J. Chem. Theory Comput.}\ } (\bibinfo {year} {2023})}\BibitemShut {NoStop}%
\bibitem [{\citenamefont {Guimar{\~{a}}es}\ \emph {et~al.}(2024)\citenamefont {Guimar{\~{a}}es}, \citenamefont {Vasilevskiy},\ and\ \citenamefont {Barbosa}}]{Guimaraes2024digitalquantum}%
  \BibitemOpen
  \bibfield  {author} {\bibinfo {author} {\bibfnamefont {J.~D.}\ \bibnamefont {Guimar{\~{a}}es}}, \bibinfo {author} {\bibfnamefont {M.~I.}\ \bibnamefont {Vasilevskiy}},\ and\ \bibinfo {author} {\bibfnamefont {L.~S.}\ \bibnamefont {Barbosa}},\ }\href {https://doi.org/10.22331/q-2024-02-05-1242} {\bibfield  {journal} {\bibinfo  {journal} {{Quantum}}\ }\textbf {\bibinfo {volume} {8}},\ \bibinfo {pages} {1242} (\bibinfo {year} {2024})}\BibitemShut {NoStop}%
\bibitem [{\citenamefont {Walters}\ and\ \citenamefont {Wang}(2024)}]{walters2024path}%
  \BibitemOpen
  \bibfield  {author} {\bibinfo {author} {\bibfnamefont {P.~L.}\ \bibnamefont {Walters}}\ and\ \bibinfo {author} {\bibfnamefont {F.}~\bibnamefont {Wang}},\ }\href@noop {} {\bibfield  {journal} {\bibinfo  {journal} {Phys.~Rev.~Research.}\ }\textbf {\bibinfo {volume} {6}},\ \bibinfo {pages} {013135} (\bibinfo {year} {2024})}\BibitemShut {NoStop}%
\bibitem [{\citenamefont {Li}\ \emph {et~al.}(2024)\citenamefont {Li}, \citenamefont {Lyu}, \citenamefont {Wang}, \citenamefont {Xu}, \citenamefont {Zheng},\ and\ \citenamefont {Yan}}]{li2024towards}%
  \BibitemOpen
  \bibfield  {author} {\bibinfo {author} {\bibfnamefont {X.}~\bibnamefont {Li}}, \bibinfo {author} {\bibfnamefont {S.-X.}\ \bibnamefont {Lyu}}, \bibinfo {author} {\bibfnamefont {Y.}~\bibnamefont {Wang}}, \bibinfo {author} {\bibfnamefont {R.-X.}\ \bibnamefont {Xu}}, \bibinfo {author} {\bibfnamefont {X.}~\bibnamefont {Zheng}},\ and\ \bibinfo {author} {\bibfnamefont {Y.}~\bibnamefont {Yan}},\ }\href@noop {} {\bibfield  {journal} {\bibinfo  {journal} {arXiv preprint arXiv:2401.17255}\ } (\bibinfo {year} {2024})}\BibitemShut {NoStop}%
\bibitem [{\citenamefont {Suess}\ \emph {et~al.}(2014)\citenamefont {Suess}, \citenamefont {Eisfeld},\ and\ \citenamefont {Strunz}}]{eisfeld2014hierarchy}%
  \BibitemOpen
  \bibfield  {author} {\bibinfo {author} {\bibfnamefont {D.}~\bibnamefont {Suess}}, \bibinfo {author} {\bibfnamefont {A.}~\bibnamefont {Eisfeld}},\ and\ \bibinfo {author} {\bibfnamefont {W.}~\bibnamefont {Strunz}},\ }\href@noop {} {\bibfield  {journal} {\bibinfo  {journal} {Phys.~Rev.~Lett.}\ }\textbf {\bibinfo {volume} {113}},\ \bibinfo {pages} {150403} (\bibinfo {year} {2014})}\BibitemShut {NoStop}%
\bibitem [{\citenamefont {Suess}\ \emph {et~al.}(2015)\citenamefont {Suess}, \citenamefont {Strunz},\ and\ \citenamefont {Eisfeld}}]{eisfeld2015hierarchical}%
  \BibitemOpen
  \bibfield  {author} {\bibinfo {author} {\bibfnamefont {D.}~\bibnamefont {Suess}}, \bibinfo {author} {\bibfnamefont {W.~T.}\ \bibnamefont {Strunz}},\ and\ \bibinfo {author} {\bibfnamefont {A.}~\bibnamefont {Eisfeld}},\ }\href@noop {} {\bibfield  {journal} {\bibinfo  {journal} {J. Stat. Phys.}\ }\textbf {\bibinfo {volume} {159}},\ \bibinfo {pages} {1408} (\bibinfo {year} {2015})}\BibitemShut {NoStop}%
\bibitem [{\citenamefont {Di{\'o}si}\ and\ \citenamefont {Strunz}(1997)}]{strunz1997pla}%
  \BibitemOpen
  \bibfield  {author} {\bibinfo {author} {\bibfnamefont {L.}~\bibnamefont {Di{\'o}si}}\ and\ \bibinfo {author} {\bibfnamefont {W.~T.}\ \bibnamefont {Strunz}},\ }\href@noop {} {\bibfield  {journal} {\bibinfo  {journal} {Phys. Lett. A}\ }\textbf {\bibinfo {volume} {235}},\ \bibinfo {pages} {569} (\bibinfo {year} {1997})}\BibitemShut {NoStop}%
\bibitem [{\citenamefont {Di\'osi}\ \emph {et~al.}(1998)\citenamefont {Di\'osi}, \citenamefont {Gisin},\ and\ \citenamefont {Strunz}}]{strunz1998pra}%
  \BibitemOpen
  \bibfield  {author} {\bibinfo {author} {\bibfnamefont {L.}~\bibnamefont {Di\'osi}}, \bibinfo {author} {\bibfnamefont {N.}~\bibnamefont {Gisin}},\ and\ \bibinfo {author} {\bibfnamefont {W.~T.}\ \bibnamefont {Strunz}},\ }\href {https://doi.org/10.1103/PhysRevA.58.1699} {\bibfield  {journal} {\bibinfo  {journal} {Phys. Rev. A}\ }\textbf {\bibinfo {volume} {58}},\ \bibinfo {pages} {1699} (\bibinfo {year} {1998})}\BibitemShut {NoStop}%
\bibitem [{\citenamefont {Strunz}\ \emph {et~al.}(1999)\citenamefont {Strunz}, \citenamefont {Di{\'o}si},\ and\ \citenamefont {Gisin}}]{strunz1999prl}%
  \BibitemOpen
  \bibfield  {author} {\bibinfo {author} {\bibfnamefont {W.~T.}\ \bibnamefont {Strunz}}, \bibinfo {author} {\bibfnamefont {L.}~\bibnamefont {Di{\'o}si}},\ and\ \bibinfo {author} {\bibfnamefont {N.}~\bibnamefont {Gisin}},\ }\href@noop {} {\bibfield  {journal} {\bibinfo  {journal} {Phys.~Rev.~Lett.}\ }\textbf {\bibinfo {volume} {82}},\ \bibinfo {pages} {1801} (\bibinfo {year} {1999})}\BibitemShut {NoStop}%
\bibitem [{\citenamefont {Yu}\ \emph {et~al.}(1999)\citenamefont {Yu}, \citenamefont {Di{\'o}si}, \citenamefont {Gisin},\ and\ \citenamefont {Strunz}}]{yu1999perturb}%
  \BibitemOpen
  \bibfield  {author} {\bibinfo {author} {\bibfnamefont {T.}~\bibnamefont {Yu}}, \bibinfo {author} {\bibfnamefont {L.}~\bibnamefont {Di{\'o}si}}, \bibinfo {author} {\bibfnamefont {N.}~\bibnamefont {Gisin}},\ and\ \bibinfo {author} {\bibfnamefont {W.~T.}\ \bibnamefont {Strunz}},\ }\href@noop {} {\bibfield  {journal} {\bibinfo  {journal} {Phys.~Rev.~A}\ }\textbf {\bibinfo {volume} {60}},\ \bibinfo {pages} {91} (\bibinfo {year} {1999})}\BibitemShut {NoStop}%
\bibitem [{\citenamefont {Jing}\ and\ \citenamefont {Yu}(2010)}]{yu2010prl}%
  \BibitemOpen
  \bibfield  {author} {\bibinfo {author} {\bibfnamefont {J.}~\bibnamefont {Jing}}\ and\ \bibinfo {author} {\bibfnamefont {T.}~\bibnamefont {Yu}},\ }\href@noop {} {\bibfield  {journal} {\bibinfo  {journal} {Phys.~Rev.~Lett.}\ }\textbf {\bibinfo {volume} {105}},\ \bibinfo {pages} {240403} (\bibinfo {year} {2010})}\BibitemShut {NoStop}%
\bibitem [{\citenamefont {Li}\ \emph {et~al.}(2014)\citenamefont {Li}, \citenamefont {Yip}, \citenamefont {Deng}, \citenamefont {Chen}, \citenamefont {Yu}, \citenamefont {You},\ and\ \citenamefont {Lam}}]{you2014nmqsd}%
  \BibitemOpen
  \bibfield  {author} {\bibinfo {author} {\bibfnamefont {Z.-Z.}\ \bibnamefont {Li}}, \bibinfo {author} {\bibfnamefont {C.-T.}\ \bibnamefont {Yip}}, \bibinfo {author} {\bibfnamefont {H.-Y.}\ \bibnamefont {Deng}}, \bibinfo {author} {\bibfnamefont {M.}~\bibnamefont {Chen}}, \bibinfo {author} {\bibfnamefont {T.}~\bibnamefont {Yu}}, \bibinfo {author} {\bibfnamefont {J.}~\bibnamefont {You}},\ and\ \bibinfo {author} {\bibfnamefont {C.-H.}\ \bibnamefont {Lam}},\ }\href@noop {} {\bibfield  {journal} {\bibinfo  {journal} {Phys.~Rev.~A}\ }\textbf {\bibinfo {volume} {90}},\ \bibinfo {pages} {022122} (\bibinfo {year} {2014})}\BibitemShut {NoStop}%
\bibitem [{\citenamefont {Luo}\ \emph {et~al.}(2015)\citenamefont {Luo}, \citenamefont {Lam}, \citenamefont {Wu}, \citenamefont {Yu}, \citenamefont {Lin},\ and\ \citenamefont {You}}]{you2015higher}%
  \BibitemOpen
  \bibfield  {author} {\bibinfo {author} {\bibfnamefont {D.-W.}\ \bibnamefont {Luo}}, \bibinfo {author} {\bibfnamefont {C.-H.}\ \bibnamefont {Lam}}, \bibinfo {author} {\bibfnamefont {L.-A.}\ \bibnamefont {Wu}}, \bibinfo {author} {\bibfnamefont {T.}~\bibnamefont {Yu}}, \bibinfo {author} {\bibfnamefont {H.-Q.}\ \bibnamefont {Lin}},\ and\ \bibinfo {author} {\bibfnamefont {J.}~\bibnamefont {You}},\ }\href@noop {} {\bibfield  {journal} {\bibinfo  {journal} {Phys.~Rev.~A}\ }\textbf {\bibinfo {volume} {92}},\ \bibinfo {pages} {022119} (\bibinfo {year} {2015})}\BibitemShut {NoStop}%
\bibitem [{\citenamefont {Gao}\ \emph {et~al.}(2022)\citenamefont {Gao}, \citenamefont {Ren}, \citenamefont {Eisfeld},\ and\ \citenamefont {Shuai}}]{gao2022non}%
  \BibitemOpen
  \bibfield  {author} {\bibinfo {author} {\bibfnamefont {X.}~\bibnamefont {Gao}}, \bibinfo {author} {\bibfnamefont {J.}~\bibnamefont {Ren}}, \bibinfo {author} {\bibfnamefont {A.}~\bibnamefont {Eisfeld}},\ and\ \bibinfo {author} {\bibfnamefont {Z.}~\bibnamefont {Shuai}},\ }\href@noop {} {\bibfield  {journal} {\bibinfo  {journal} {Phys.~Rev.~A}\ }\textbf {\bibinfo {volume} {105}},\ \bibinfo {pages} {L030202} (\bibinfo {year} {2022})}\BibitemShut {NoStop}%
\bibitem [{\citenamefont {Flannigan}\ \emph {et~al.}(2022)\citenamefont {Flannigan}, \citenamefont {Damanet},\ and\ \citenamefont {Daley}}]{flannigan2022many}%
  \BibitemOpen
  \bibfield  {author} {\bibinfo {author} {\bibfnamefont {S.}~\bibnamefont {Flannigan}}, \bibinfo {author} {\bibfnamefont {F.}~\bibnamefont {Damanet}},\ and\ \bibinfo {author} {\bibfnamefont {A.~J.}\ \bibnamefont {Daley}},\ }\href@noop {} {\bibfield  {journal} {\bibinfo  {journal} {Phys.~Rev.~Lett.}\ }\textbf {\bibinfo {volume} {128}},\ \bibinfo {pages} {063601} (\bibinfo {year} {2022})}\BibitemShut {NoStop}%
\bibitem [{\citenamefont {Moroder}\ \emph {et~al.}(2023)\citenamefont {Moroder}, \citenamefont {Grundner}, \citenamefont {Damanet}, \citenamefont {Schollw{\"o}ck}, \citenamefont {Mardazad}, \citenamefont {Flannigan}, \citenamefont {K{\"o}hler},\ and\ \citenamefont {Paeckel}}]{moroder2023stable}%
  \BibitemOpen
  \bibfield  {author} {\bibinfo {author} {\bibfnamefont {M.}~\bibnamefont {Moroder}}, \bibinfo {author} {\bibfnamefont {M.}~\bibnamefont {Grundner}}, \bibinfo {author} {\bibfnamefont {F.}~\bibnamefont {Damanet}}, \bibinfo {author} {\bibfnamefont {U.}~\bibnamefont {Schollw{\"o}ck}}, \bibinfo {author} {\bibfnamefont {S.}~\bibnamefont {Mardazad}}, \bibinfo {author} {\bibfnamefont {S.}~\bibnamefont {Flannigan}}, \bibinfo {author} {\bibfnamefont {T.}~\bibnamefont {K{\"o}hler}},\ and\ \bibinfo {author} {\bibfnamefont {S.}~\bibnamefont {Paeckel}},\ }\href@noop {} {\bibfield  {journal} {\bibinfo  {journal} {Phys.~Rev.~B}\ }\textbf {\bibinfo {volume} {107}},\ \bibinfo {pages} {214310} (\bibinfo {year} {2023})}\BibitemShut {NoStop}%
\bibitem [{\citenamefont {Yuan}\ \emph {et~al.}(2019)\citenamefont {Yuan}, \citenamefont {Endo}, \citenamefont {Zhao}, \citenamefont {Li},\ and\ \citenamefont {Benjamin}}]{yuan2019theory}%
  \BibitemOpen
  \bibfield  {author} {\bibinfo {author} {\bibfnamefont {X.}~\bibnamefont {Yuan}}, \bibinfo {author} {\bibfnamefont {S.}~\bibnamefont {Endo}}, \bibinfo {author} {\bibfnamefont {Q.}~\bibnamefont {Zhao}}, \bibinfo {author} {\bibfnamefont {Y.}~\bibnamefont {Li}},\ and\ \bibinfo {author} {\bibfnamefont {S.~C.}\ \bibnamefont {Benjamin}},\ }\href@noop {} {\bibfield  {journal} {\bibinfo  {journal} {Quantum}\ }\textbf {\bibinfo {volume} {3}},\ \bibinfo {pages} {191} (\bibinfo {year} {2019})}\BibitemShut {NoStop}%
\bibitem [{\citenamefont {Endo}\ \emph {et~al.}(2020)\citenamefont {Endo}, \citenamefont {Sun}, \citenamefont {Li}, \citenamefont {Benjamin},\ and\ \citenamefont {Yuan}}]{endo2020variational}%
  \BibitemOpen
  \bibfield  {author} {\bibinfo {author} {\bibfnamefont {S.}~\bibnamefont {Endo}}, \bibinfo {author} {\bibfnamefont {J.}~\bibnamefont {Sun}}, \bibinfo {author} {\bibfnamefont {Y.}~\bibnamefont {Li}}, \bibinfo {author} {\bibfnamefont {S.~C.}\ \bibnamefont {Benjamin}},\ and\ \bibinfo {author} {\bibfnamefont {X.}~\bibnamefont {Yuan}},\ }\href@noop {} {\bibfield  {journal} {\bibinfo  {journal} {Phys.~Rev.~Lett.}\ }\textbf {\bibinfo {volume} {125}},\ \bibinfo {pages} {010501} (\bibinfo {year} {2020})}\BibitemShut {NoStop}%
\bibitem [{\citenamefont {Schlimgen}\ \emph {et~al.}(2021)\citenamefont {Schlimgen}, \citenamefont {Head-Marsden}, \citenamefont {Sager}, \citenamefont {Narang},\ and\ \citenamefont {Mazziotti}}]{schlimgen2021quantum}%
  \BibitemOpen
  \bibfield  {author} {\bibinfo {author} {\bibfnamefont {A.~W.}\ \bibnamefont {Schlimgen}}, \bibinfo {author} {\bibfnamefont {K.}~\bibnamefont {Head-Marsden}}, \bibinfo {author} {\bibfnamefont {L.~M.}\ \bibnamefont {Sager}}, \bibinfo {author} {\bibfnamefont {P.}~\bibnamefont {Narang}},\ and\ \bibinfo {author} {\bibfnamefont {D.~A.}\ \bibnamefont {Mazziotti}},\ }\href@noop {} {\bibfield  {journal} {\bibinfo  {journal} {Phys.~Rev.~Lett.}\ }\textbf {\bibinfo {volume} {127}},\ \bibinfo {pages} {270503} (\bibinfo {year} {2021})}\BibitemShut {NoStop}%
\bibitem [{Note1()}]{Note1}%
  \BibitemOpen
  \bibinfo {note} {With our choice of the correlation functions of the stochastic processes we follow the one of the original NMQSD derivation \cite {strunz1997pla,strunz1998pra}. There exist other choices fore the noise-correlations which might give numerical advances e.g., for high temperature~\cite {zhao2016hierarchy,shi2016alternative}.}\BibitemShut {Stop}%
\bibitem [{\citenamefont {Okuma}\ and\ \citenamefont {Nakagawa}(2022)}]{c1}%
  \BibitemOpen
  \bibfield  {author} {\bibinfo {author} {\bibfnamefont {N.}~\bibnamefont {Okuma}}\ and\ \bibinfo {author} {\bibfnamefont {Y.~O.}\ \bibnamefont {Nakagawa}},\ }\href@noop {} {\bibfield  {journal} {\bibinfo  {journal} {Phys.~Rev.~B}\ }\textbf {\bibinfo {volume} {105}},\ \bibinfo {pages} {054304} (\bibinfo {year} {2022})}\BibitemShut {NoStop}%
\bibitem [{\citenamefont {Heyl}(2014)}]{heyl2014dynamical}%
  \BibitemOpen
  \bibfield  {author} {\bibinfo {author} {\bibfnamefont {M.}~\bibnamefont {Heyl}},\ }\href@noop {} {\bibfield  {journal} {\bibinfo  {journal} {Phys.~Rev.~Lett.}\ }\textbf {\bibinfo {volume} {113}},\ \bibinfo {pages} {205701} (\bibinfo {year} {2014})}\BibitemShut {NoStop}%
\bibitem [{\citenamefont {Zvyagin}(2016)}]{zvyagin2016dynamical}%
  \BibitemOpen
  \bibfield  {author} {\bibinfo {author} {\bibfnamefont {A.}~\bibnamefont {Zvyagin}},\ }\href@noop {} {\bibfield  {journal} {\bibinfo  {journal} {Low Temp. Phys.}\ }\textbf {\bibinfo {volume} {42}},\ \bibinfo {pages} {971} (\bibinfo {year} {2016})}\BibitemShut {NoStop}%
\bibitem [{\citenamefont {Heyl}\ and\ \citenamefont {Budich}(2017)}]{heyl2017dynamical}%
  \BibitemOpen
  \bibfield  {author} {\bibinfo {author} {\bibfnamefont {M.}~\bibnamefont {Heyl}}\ and\ \bibinfo {author} {\bibfnamefont {J.}~\bibnamefont {Budich}},\ }\href@noop {} {\bibfield  {journal} {\bibinfo  {journal} {Phys.~Rev.~B}\ }\textbf {\bibinfo {volume} {96}},\ \bibinfo {pages} {180304} (\bibinfo {year} {2017})}\BibitemShut {NoStop}%
\bibitem [{\citenamefont {Heyl}(2019)}]{heyl2019dynamical}%
  \BibitemOpen
  \bibfield  {author} {\bibinfo {author} {\bibfnamefont {M.}~\bibnamefont {Heyl}},\ }\href@noop {} {\bibfield  {journal} {\bibinfo  {journal} {Europhys. Lett.}\ }\textbf {\bibinfo {volume} {125}},\ \bibinfo {pages} {26001} (\bibinfo {year} {2019})}\BibitemShut {NoStop}%
\bibitem [{\citenamefont {Puebla}(2020)}]{puebla2020finite}%
  \BibitemOpen
  \bibfield  {author} {\bibinfo {author} {\bibfnamefont {R.}~\bibnamefont {Puebla}},\ }\href@noop {} {\bibfield  {journal} {\bibinfo  {journal} {Phys.~Rev.~B}\ }\textbf {\bibinfo {volume} {102}},\ \bibinfo {pages} {220302} (\bibinfo {year} {2020})}\BibitemShut {NoStop}%
\bibitem [{\citenamefont {Norambuena}\ \emph {et~al.}(2020)\citenamefont {Norambuena}, \citenamefont {Tancara},\ and\ \citenamefont {Coto}}]{norambuena2020coding}%
  \BibitemOpen
  \bibfield  {author} {\bibinfo {author} {\bibfnamefont {A.}~\bibnamefont {Norambuena}}, \bibinfo {author} {\bibfnamefont {D.}~\bibnamefont {Tancara}},\ and\ \bibinfo {author} {\bibfnamefont {R.}~\bibnamefont {Coto}},\ }\href@noop {} {\bibfield  {journal} {\bibinfo  {journal} {Eur. J. Phys.}\ }\textbf {\bibinfo {volume} {41}},\ \bibinfo {pages} {045404} (\bibinfo {year} {2020})}\BibitemShut {NoStop}%
\bibitem [{\citenamefont {Dolgitzer}\ \emph {et~al.}(2021)\citenamefont {Dolgitzer}, \citenamefont {Zeng},\ and\ \citenamefont {Chen}}]{dolgitzer2021dynamical}%
  \BibitemOpen
  \bibfield  {author} {\bibinfo {author} {\bibfnamefont {D.}~\bibnamefont {Dolgitzer}}, \bibinfo {author} {\bibfnamefont {D.}~\bibnamefont {Zeng}},\ and\ \bibinfo {author} {\bibfnamefont {Y.}~\bibnamefont {Chen}},\ }\href@noop {} {\bibfield  {journal} {\bibinfo  {journal} {Opt. Express.}\ }\textbf {\bibinfo {volume} {29}},\ \bibinfo {pages} {23988} (\bibinfo {year} {2021})}\BibitemShut {NoStop}%
\bibitem [{\citenamefont {Gao}\ and\ \citenamefont {Eisfeld}(2019)}]{gao2019charge}%
  \BibitemOpen
  \bibfield  {author} {\bibinfo {author} {\bibfnamefont {X.}~\bibnamefont {Gao}}\ and\ \bibinfo {author} {\bibfnamefont {A.}~\bibnamefont {Eisfeld}},\ }\href@noop {} {\bibfield  {journal} {\bibinfo  {journal} {J.~Chem.~Phys.}\ }\textbf {\bibinfo {volume} {150}} (\bibinfo {year} {2019})}\BibitemShut {NoStop}%
\bibitem [{\citenamefont {Somma}\ \emph {et~al.}(2003)\citenamefont {Somma}, \citenamefont {Ortiz}, \citenamefont {Knill},\ and\ \citenamefont {Gubernatis}}]{somma2003quantum}%
  \BibitemOpen
  \bibfield  {author} {\bibinfo {author} {\bibfnamefont {R.}~\bibnamefont {Somma}}, \bibinfo {author} {\bibfnamefont {G.}~\bibnamefont {Ortiz}}, \bibinfo {author} {\bibfnamefont {E.}~\bibnamefont {Knill}},\ and\ \bibinfo {author} {\bibfnamefont {J.}~\bibnamefont {Gubernatis}},\ }\href@noop {} {\bibfield  {journal} {\bibinfo  {journal} {Int. J. Quantum Inf.}\ }\textbf {\bibinfo {volume} {1}},\ \bibinfo {pages} {189} (\bibinfo {year} {2003})}\BibitemShut {NoStop}%
\bibitem [{\citenamefont {Sawaya}\ \emph {et~al.}(2020)\citenamefont {Sawaya}, \citenamefont {Menke}, \citenamefont {Kyaw}, \citenamefont {Johri}, \citenamefont {Aspuru-Guzik},\ and\ \citenamefont {Guerreschi}}]{sawaya2020resource}%
  \BibitemOpen
  \bibfield  {author} {\bibinfo {author} {\bibfnamefont {N.~P.}\ \bibnamefont {Sawaya}}, \bibinfo {author} {\bibfnamefont {T.}~\bibnamefont {Menke}}, \bibinfo {author} {\bibfnamefont {T.~H.}\ \bibnamefont {Kyaw}}, \bibinfo {author} {\bibfnamefont {S.}~\bibnamefont {Johri}}, \bibinfo {author} {\bibfnamefont {A.}~\bibnamefont {Aspuru-Guzik}},\ and\ \bibinfo {author} {\bibfnamefont {G.~G.}\ \bibnamefont {Guerreschi}},\ }\href@noop {} {\bibfield  {journal} {\bibinfo  {journal} {npj Quantum Inf.}\ }\textbf {\bibinfo {volume} {6}},\ \bibinfo {pages} {49} (\bibinfo {year} {2020})}\BibitemShut {NoStop}%
\bibitem [{\citenamefont {Di~Matteo}\ \emph {et~al.}(2021)\citenamefont {Di~Matteo}, \citenamefont {McCoy}, \citenamefont {Gysbers}, \citenamefont {Miyagi}, \citenamefont {Woloshyn},\ and\ \citenamefont {Navr{\'a}til}}]{di2021improving}%
  \BibitemOpen
  \bibfield  {author} {\bibinfo {author} {\bibfnamefont {O.}~\bibnamefont {Di~Matteo}}, \bibinfo {author} {\bibfnamefont {A.}~\bibnamefont {McCoy}}, \bibinfo {author} {\bibfnamefont {P.}~\bibnamefont {Gysbers}}, \bibinfo {author} {\bibfnamefont {T.}~\bibnamefont {Miyagi}}, \bibinfo {author} {\bibfnamefont {R.}~\bibnamefont {Woloshyn}},\ and\ \bibinfo {author} {\bibfnamefont {P.}~\bibnamefont {Navr{\'a}til}},\ }\href@noop {} {\bibfield  {journal} {\bibinfo  {journal} {Phys.~Rev.~A}\ }\textbf {\bibinfo {volume} {103}},\ \bibinfo {pages} {042405} (\bibinfo {year} {2021})}\BibitemShut {NoStop}%
\bibitem [{\citenamefont {Johansson}\ \emph {et~al.}(2012)\citenamefont {Johansson}, \citenamefont {Nation},\ and\ \citenamefont {Nori}}]{johansson2012qutip}%
  \BibitemOpen
  \bibfield  {author} {\bibinfo {author} {\bibfnamefont {J.~R.}\ \bibnamefont {Johansson}}, \bibinfo {author} {\bibfnamefont {P.~D.}\ \bibnamefont {Nation}},\ and\ \bibinfo {author} {\bibfnamefont {F.}~\bibnamefont {Nori}},\ }\href@noop {} {\bibfield  {journal} {\bibinfo  {journal} {Comp. Phys. Comm.}\ }\textbf {\bibinfo {volume} {183}},\ \bibinfo {pages} {1760} (\bibinfo {year} {2012})}\BibitemShut {NoStop}%
\bibitem [{\citenamefont {Johansson}\ \emph {et~al.}(2013)\citenamefont {Johansson}, \citenamefont {Nation},\ and\ \citenamefont {Nori}}]{JOHANSSON2013qutip}%
  \BibitemOpen
  \bibfield  {author} {\bibinfo {author} {\bibfnamefont {J.}~\bibnamefont {Johansson}}, \bibinfo {author} {\bibfnamefont {P.}~\bibnamefont {Nation}},\ and\ \bibinfo {author} {\bibfnamefont {F.}~\bibnamefont {Nori}},\ }\href@noop {} {\bibfield  {journal} {\bibinfo  {journal} {Comp. Phys. Comm.}\ }\textbf {\bibinfo {volume} {184}},\ \bibinfo {pages} {1234} (\bibinfo {year} {2013})}\BibitemShut {NoStop}%
\bibitem [{\citenamefont {Schlimgen}\ \emph {et~al.}(2022)\citenamefont {Schlimgen}, \citenamefont {Head-Marsden}, \citenamefont {Sager}, \citenamefont {Narang},\ and\ \citenamefont {Mazziotti}}]{qs_unitary_decomposition}%
  \BibitemOpen
  \bibfield  {author} {\bibinfo {author} {\bibfnamefont {A.~W.}\ \bibnamefont {Schlimgen}}, \bibinfo {author} {\bibfnamefont {K.}~\bibnamefont {Head-Marsden}}, \bibinfo {author} {\bibfnamefont {L.~M.}\ \bibnamefont {Sager}}, \bibinfo {author} {\bibfnamefont {P.}~\bibnamefont {Narang}},\ and\ \bibinfo {author} {\bibfnamefont {D.~A.}\ \bibnamefont {Mazziotti}},\ }\href@noop {} {\bibfield  {journal} {\bibinfo  {journal} {Phys.~Rev.~Research.}\ }\textbf {\bibinfo {volume} {4}},\ \bibinfo {pages} {023216} (\bibinfo {year} {2022})}\BibitemShut {NoStop}%
\bibitem [{\citenamefont {Li}\ \emph {et~al.}(2023)\citenamefont {Li}, \citenamefont {Ren}, \citenamefont {Huai}, \citenamefont {Cai}, \citenamefont {Shuai},\ and\ \citenamefont {Zhang}}]{li2023efficient}%
  \BibitemOpen
  \bibfield  {author} {\bibinfo {author} {\bibfnamefont {W.}~\bibnamefont {Li}}, \bibinfo {author} {\bibfnamefont {J.}~\bibnamefont {Ren}}, \bibinfo {author} {\bibfnamefont {S.}~\bibnamefont {Huai}}, \bibinfo {author} {\bibfnamefont {T.}~\bibnamefont {Cai}}, \bibinfo {author} {\bibfnamefont {Z.}~\bibnamefont {Shuai}},\ and\ \bibinfo {author} {\bibfnamefont {S.}~\bibnamefont {Zhang}},\ }\href@noop {} {\bibfield  {journal} {\bibinfo  {journal} {Phys.~Rev.~Research.}\ }\textbf {\bibinfo {volume} {5}},\ \bibinfo {pages} {023046} (\bibinfo {year} {2023})}\BibitemShut {NoStop}%
\bibitem [{\citenamefont {Ikeda}\ and\ \citenamefont {Scholes}(2020)}]{ikeda2020generalization}%
  \BibitemOpen
  \bibfield  {author} {\bibinfo {author} {\bibfnamefont {T.}~\bibnamefont {Ikeda}}\ and\ \bibinfo {author} {\bibfnamefont {G.~D.}\ \bibnamefont {Scholes}},\ }\href@noop {} {\bibfield  {journal} {\bibinfo  {journal} {J.~Chem.~Phys.}\ }\textbf {\bibinfo {volume} {152}} (\bibinfo {year} {2020})}\BibitemShut {NoStop}%
\bibitem [{\citenamefont {Ke}\ and\ \citenamefont {Zhao}(2016)}]{zhao2016hierarchy}%
  \BibitemOpen
  \bibfield  {author} {\bibinfo {author} {\bibfnamefont {Y.}~\bibnamefont {Ke}}\ and\ \bibinfo {author} {\bibfnamefont {Y.}~\bibnamefont {Zhao}},\ }\href@noop {} {\bibfield  {journal} {\bibinfo  {journal} {J.~Chem.~Phys.}\ }\textbf {\bibinfo {volume} {145}},\ \bibinfo {pages} {024101} (\bibinfo {year} {2016})}\BibitemShut {NoStop}%
\bibitem [{\citenamefont {Song}\ \emph {et~al.}(2016)\citenamefont {Song}, \citenamefont {Song},\ and\ \citenamefont {Shi}}]{shi2016alternative}%
  \BibitemOpen
  \bibfield  {author} {\bibinfo {author} {\bibfnamefont {K.}~\bibnamefont {Song}}, \bibinfo {author} {\bibfnamefont {L.}~\bibnamefont {Song}},\ and\ \bibinfo {author} {\bibfnamefont {Q.}~\bibnamefont {Shi}},\ }\href@noop {} {\bibfield  {journal} {\bibinfo  {journal} {J.~Chem.~Phys.}\ }\textbf {\bibinfo {volume} {144}},\ \bibinfo {pages} {224105} (\bibinfo {year} {2016})}\BibitemShut {NoStop}%
\end{thebibliography}%
	
\end{document}